\documentclass[sn-mathphys,Numbered]{sn-jnl}

\usepackage{graphicx}%
\usepackage{multirow}%
\usepackage{amsmath,amssymb,amsfonts}%
\usepackage{amsthm}%
\usepackage{mathrsfs}%
\usepackage[title]{appendix}%
\usepackage{xcolor}%
\usepackage{textcomp}%
\usepackage{manyfoot}%
\usepackage{booktabs}%
\usepackage{algorithm}%
\usepackage{algorithmicx}%
\usepackage{algpseudocode}%
\usepackage{listings}%

\theoremstyle{thmstyleone}%
\theoremstyle{thmstyletwo}%

\theoremstyle{thmstylethree}%

\begin{document}

\title[Does energy always have mass?]{Mass-energy equivalence and the
  gravitational redshift: Does energy always have mass?}


\author*[1]{\fnm{Germano} \sur{D'Abramo}}\email{germano.dabramo@gmail.com}
\equalcont{ORCID: 0000-0003-1277-7418}



\affil*[1]{\orgname{Ministero dell'Istruzione, dell'Universit\`a e della
    Ricerca}, \orgaddress{\postcode{00041},
    \state{Albano Laziale}, \country{Italy}}}




\abstract{One of the most widespread interpretations of the mass-energy
  equivalence establishes that not only can mass be transformed into energy
  (e.g., through nuclear fission, fusion, or annihilation) but that every type
  of energy also has mass (via the mass-energy equivalence formula). Here, we
  show that this is not always the case. With the help a few thought
  experiments, we show that, for instance, the electric potential energy of a
  charged capacitor should not contribute to the capacitor’s gravitational rest
  mass (while still contributing to its linear momentum). That result is in
  agreement with the fact that light (ultimately, an electromagnetic
  phenomenon) has momentum but not rest mass.}





\maketitle

\section{Introduction}
\label{se1}

The actual meaning and correct interpretation of the celebrated mass-energy
equivalence $E = mc^2$ is still a matter of discussion among scholars.
For a far-from-complete collection of references to the existing literature on
mass-energy equivalence derivation, discussion, and interpretation, see, for
instance, the study of Einstein~\cite{e05b}, Planck~\cite{p07}, Laue~\cite{lau},
Klein~\cite{kle}, Einstein~\cite{e46}, Ives Herbert~\cite{i52},
Jammer~\cite{j61}, Stachel and Torretti~\cite{st82}, Rohrlich~\cite{roh},
Ohanian~\cite{o08,o08b}, Hecht~\cite{h11}, D'Abramo~\cite{dab3},
Stanford Encyclopedia of Physics~\cite{sta}, and references therein.

\begin{figure}[t]
\begin{center}
\includegraphics[width=4cm]{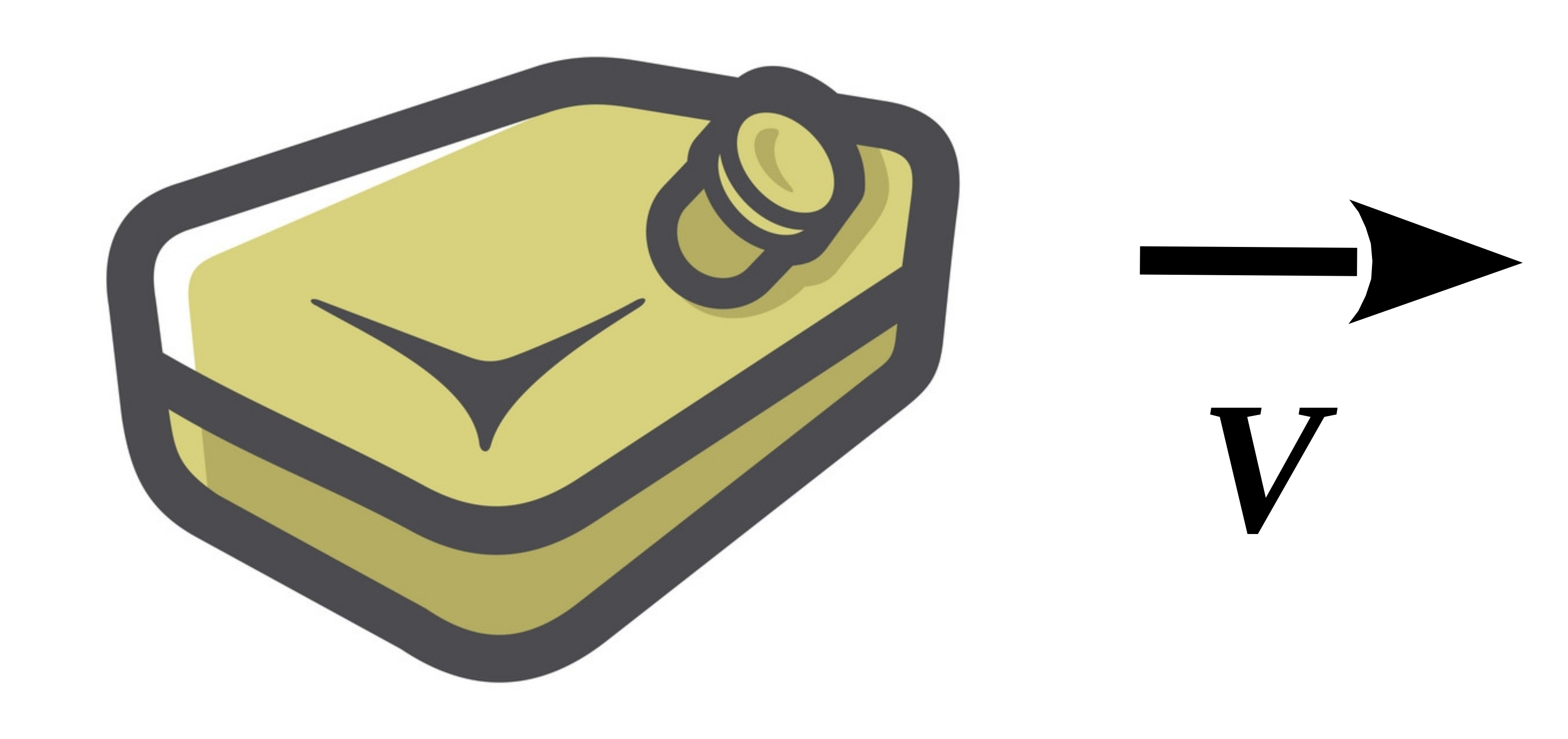}
\end{center}
\caption{Does a gasoline tank in motion have more internal (chemical) energy
  than a stationary one? That appears to be a necessary consequence of the
  crucial assumption made by Einstein in his 1905 derivation of mass-energy
  equivalence~\cite{dab}.}
\label{fig1}
\end{figure}

In a recently published paper~\cite{dab}, we reexamined Einstein's 1905
derivation of mass-energy equivalence~\cite{e05b}.
Einstein's original approach consisted in studying, in different
reference frames, the energy balance of a body emitting electromagnetic
radiation. In that paper, we showed that an unsupported assumption
stands behind the validity of Einstein's celebrated result, namely that the
motion of the body, in the form of its kinetic energy K relative to a
stationary observer $O$, does contribute to the increase in the `internal
reservoir' of energy from which the electromagnetic emission originates with
respect to $O$. We pointed out that with electromagnetic emissions or with
any non-mechanical process, the consequences implied by that assumption are not
unproblematic. As a matter of fact, in cases like those, it is much like
taking for granted that, for instance, the kinetic energy of an electric battery
in motion relative to us can contribute, for us, to the increase in the
electrical energy content of that battery. Or that the kinetic energy of a car
in motion relative to us can contribute, for us, to the increase in the energy
content of the gasoline and, ultimately, to the increase in the gasoline
mass (see Fig.~\ref{fig1}).

Moreover, in the same paper, we gave strong evidence that the mentioned
Einstein's assumption is logically equivalent, although not in a trivial way,
to assuming mass-energy equivalence from the outset. We concluded that
Einstein's original result was not {\em proving} that mass and energy are
equivalent but, more correctly, that {\em if} mass transforms into energy, it
does it according to the relation $E=mc^2$.

Furthermore, inspired by the abovementioned results, we ended up asking
whether energy always has mass. To be precise, if and when mass transforms
into energy, like, for instance, in nuclear reactions (fission, fusion,
annihilation, etc.), mass and energy are indeed related according to the
equation $E=mc^2$. However, the question is whether every form of
energy (heat, electric or gravitational potential energy, etc.) always
has an inertial/gravitational mass. 

At the end of our study~\cite{dab}, we questioned that indiscriminate
energy-to-mass conversion belief by analyzing and revising the following
thought experiment by Misner, Thorne, and Wheeler on the gravitational
frequency shift derived from the conservation of energy:\cite{mtw}

\begin{quote}
  {\small That a photon must be affected by a gravitational field Einstein
    (1911) showed from the law of conservation of energy, applied in the context
    of Newtonian gravitation theory. Let a particle of rest mass $m$ start from
    rest in a gravitational field $g$ at point ${\cal{A}}$ and fall freely for
    a distance $h$ to point ${\cal{B}}$. It gains kinetic energy $mgh$. Its
    total energy, including rest mass, becomes

    $$ m+mgh.$$

    Now, let the particle undergo an annihilation at ${\cal{B}}$,
    converting its total rest mass plus kinetic energy into a photon of
    the same energy. Let this photon travel upward in the
    gravitational field to ${\cal{A}}$. If it does not interact with gravity,
    it will have its original energy on arrival at ${\cal{A}}$. At this point 
    it could be converted by a suitable apparatus into another particle of rest
    mass $m$ (which could then repeat the whole process) plus an excess energy
    $mgh$ that costs nothing to produce. To avoid this contradiction of the
    principal [{\em sic}] of conservation of energy, which can also be stated
    in purely classical terms, Einstein saw that the photon must suffer a
    red shift. [{\em The speed of light is set as} $c=1$]
   }    
\end{quote}  

Unfortunately, Misner, Thorne, and Wheeler's argument appears to be problematic.
If a particle of rest mass $m$ starts from rest in a gravitational field $g$
at point ${\cal{A}}$ and falls freely for a distance $h$ to point ${\cal{B}}$,
that particle possesses also an energy equal to $mgh$
already at point ${\cal{A}}$. It is called gravitational potential energy.
Therefore, {\em owing to the complete mass-energy equivalence}, at point
${\cal{A}}$, that particle already has a total mass/energy equal to $m+mgh$.
It can be shown that, in a uniform gravitational field $g$, the mass $m_h$ of
a particle at height $h$ is $m_h=me^{\frac{gh}{c^2}}$, where $m$ is the proper
mass at height taken as zero. The total energy $E_{tot}$, proper mass plus
gravitational potential energy, at height $h$ is given by
$E_{tot}=mc^2e^{\frac{gh}{c^2}}$. For small distances $h$, we have
$m_h \simeq m + \frac{mgh}{c^2}$ and $E_{tot}\simeq mc^2 + mgh$. By assuming
$c=1$, like in Misner {\em et al.}~\cite{mtw}, we have that the mass and the
total energy of the particle at the height $h$ (point ${\cal{A}}$ in Misner
{\em et al}.~\cite{mtw}) are $m + mgh$. Now, if the energy of the photon
generated in the particle annihilation and traveling upward does not have its
original value on arrival at ${\cal{A}}$ (i.e., $m+mgh$), the mass of the
particle created by the suitable apparatus at the end of the process would not
have the same mass as the original particle (again, $m+mgh$), and the total
energy/mass would not be conserved. When Misner, Thorne, and Wheeler say that
the particle ``gains kinetic energy $mgh$''  on arrival at point ${\cal{B}}$,
and ``its total energy, including rest mass, becomes  $m+mgh$'', they seem to
forget that the particle already has a gravitational potential energy $mgh$,
and total energy $m+mgh$, just before starting to fall. That is demanded by the
principle of conservation of energy.

Therefore, the widely-held assumption that every energy always has mass is at
odds with the conservation of energy and the existence of the gravitational
frequency shift taken together. The thought experiment by Misner, Thorne, and
Wheeler pits the above three assumptions one against the other. They cannot be
simultaneously true. However, we concluded our paper by saying that
it is still not clear which one, among the three, is actually at
fault.\cite{dab} 
The only exception we felt like making was for the conservation of energy.

The present paper aims to clarify that issue. First, by applying energy
and linear momentum conservation, we prove that, in the case of Misner, Thorne,
and Wheeler's derivation, the gravitational potential energy of a body does, in
fact, have mass and does contribute to the total mass of the body
(Section~\ref{se2}). Within that proof, we also show that the gravitational
frequency shift is incompatible with the conservation of linear momentum.
Therefore, returning to the conclusion of the paper~\cite{dab}, the
culprit seems to be the soundness of the gravitational frequency shift
phenomenon. 

In Section~\ref{se3}, we provide a different proof showing that the
gravitational frequency shift, taken alone, is incompatible with energy
conservation. That proof does not require the assumption of complete
mass-energy equivalence. In particular, we do not even need to assume that the
gravitational potential energy of a body contributes to the total mass of the
body, as we have done in our revision of Misner, Thorne, and Wheeler's
derivation.

Finally, in Section~\ref{se4}, by using the same type of thought experiment
given in Section~\ref{se3}, we prove that energy does not always have mass.
Specifically, we analyze the case of the energy stored in a charged capacitor.
We show that the electric potential energy of a charged capacitor does not
contribute to the capacitor's rest mass while still contributing to its
momentum.

In the concluding section, we briefly summarize the results achieved in this
paper.

\section{Gravitational frequency shift and linear momentum conservation}
\label{se2}

\begin{figure*}[t]
\begin{center}
\centerline{\includegraphics[width=14cm]{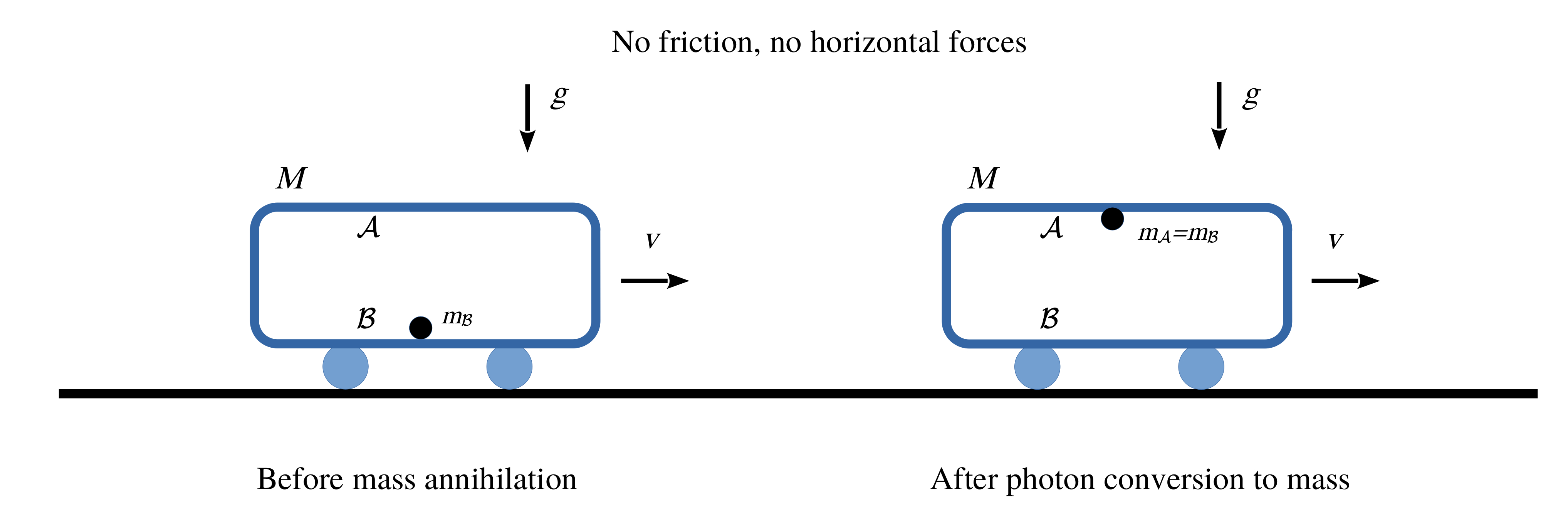}}
\end{center}
\caption{Pictorial representation of the thought experiment described in
  Section~\ref{se2}.}
\label{fig2}
\end{figure*}

Here, we show that the gravitational potential energy of a body contributes to
the total mass of the body, as assumed in our analysis of Misner, Thorne,
and Wheeler's derivation in Section~\ref{se1}.
Consider the following ideal experiment. A closed wagon of mass $M$ moves
horizontally without friction in a vertical uniform gravitational field $g$ at a
constant velocity $v$ (see Fig.~\ref{fig2}).
Inside the wagon, attached to floor ${\cal B}$, there is a particle of mass
$m_{\cal B}$. At a certain point, mass $m_{\cal B}$ annihilates into a photon of
energy $h\nu_{\cal B}=m_{\cal B}c^2$, where $h$ is the Planck constant and
$\nu_{\cal B}$ the frequency of the photon generated at point ${\cal B}$.
Then, the photon travels upward toward ceiling ${\cal A}$ and is absorbed and
converted by a suitable apparatus into another particle of mass $m_{\cal A}$.
This particle also ends up stuck to the wagon frame. The whole process happens
exclusively inside the closed wagon.
Owing to the conservation of energy, we must have that
$h\nu_{\cal B}=m_{\cal A}c^2+m_{\cal A}gh$, but according to the common
understanding, the total mass of the generated particle at point ${\cal A}$
does not include the equivalent mass of its gravitational potential energy
$m_{\cal A}gh/c^2$.

In reality, the total mass of the particle generated at point ${\cal A}$ must
be $m_{\cal A}+m_{\cal A}gh/c^2=h\nu_{\cal B}/c^2= m_{\cal B}$, and therefore, it must
include the equivalent mass of its own gravitational potential energy. Any
different scenario seems to violate the conservation of (the horizontal) linear
momentum of the closed system wagon+particle.
No horizontal external forces act upon the system, and no mass is ejected.
Therefore, the total velocity $v$ must be the same before and after the
whole process.
However, before the annihilation, the total horizontal linear momentum is
$P_i=(M+m_{\cal B})v$ while, after the conversion of the photon energy into mass
and if the total mass of the particle generated at point ${\cal A}$ is less
than $m_{\cal B}$, the total horizontal linear momentum becomes
$P_f=(M+m_{\cal A})v < P_i$. That is quite bizarre. On the other hand, by
imposing the conservation of the horizontal linear momentum, we would have an
equally strange consequence.
Without any horizontal external force acting upon the wagon and without any
mass ejection, we would see the wagon increase its velocity by itself at the
end of the whole process.

Incidentally, the above argument suggests that there is a problem with the
gravitational redshift: if the total mass of the particle generated
at point ${\cal A}$ is still $m_{\cal B}$, the energy of the photon from which it
derives is $m_{\cal B}c^2=h\nu_{\cal B}$, namely, the frequency of the photon at
point ${\cal A}$ must be the same as that at point ${\cal B}$, $\nu_A=\nu_B$.

\section{Gravitational frequency shift and the conservation of energy}
\label{se3}

Here, we give a different proof that photon (radiation) energy is not
affected by a gravitational field. In the following thought experiment, the
assumption of complete mass-energy equivalence is not used. In particular, we
do not even need to assume that the gravitational potential energy of a body
contributes to the total mass of the body as we have done in our revision of
Misner, Thorne, and Wheeler's derivation. This thought experiment has already
been applied to sound waves to show that they can escape any gravity
well~\cite{dab2}.

\begin{figure}[t]
\begin{center}
\includegraphics[width=14cm]{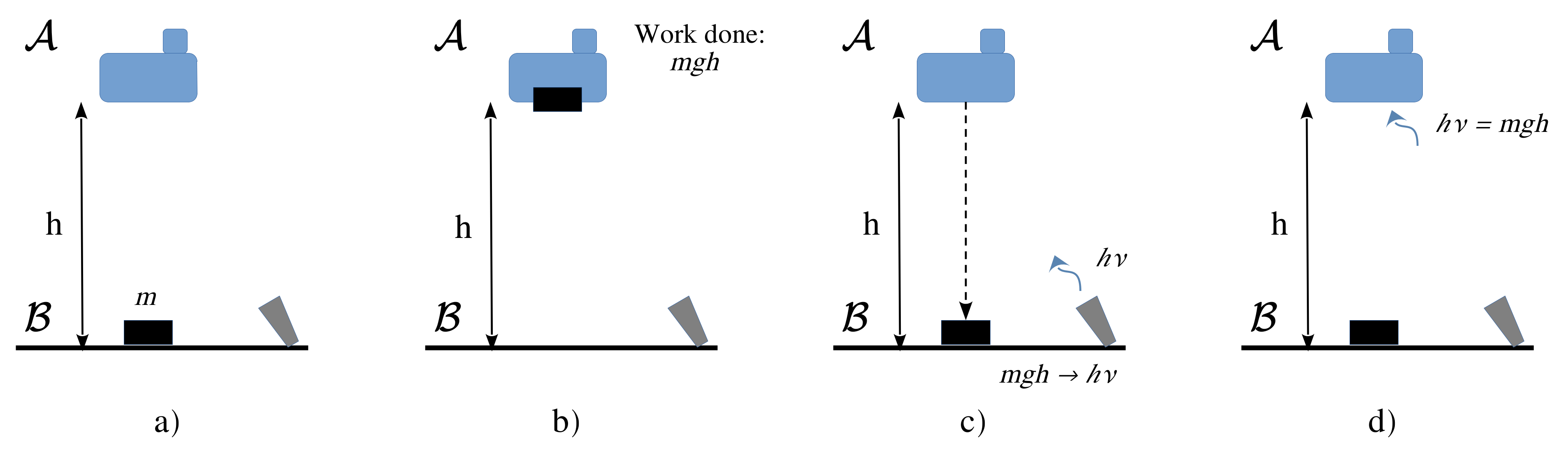}
\end{center}
\caption{Pictorial representation of the thought experiment described in
  Section~\ref{se3}.}
\label{fig3}
\end{figure}

Consider a body of mass $m$ stationary at point ${\cal B}$ and a
macroscopic apparatus stationary at point ${\cal A}$, at a height $h$ above
point ${\cal B}$ in a uniform gravitational field $g$ (Fig.~\ref{fig3}). Let
the apparatus perform mechanical work on body $m$, raising it to point
${\cal A}$.
The work done by the apparatus is equal to $mgh$, which is also equal to the
gravitational potential energy of the body $m$ relative to point ${\cal B}$.
Now, if the mass is lowered back to point ${\cal B}$ and its potential energy
conventionally (and entirely) converted into electrical energy and then
into a single photon of energy $mgh$ (ultimately emitted by a beacon), the
energy of the photon  must always be the same while climbing up the
gravitational field back to point ${\cal A}$.
The photon energy at point ${\cal A}$ must still be equal to $mgh$.
That is demanded by the conservation of energy. Through photon absorption,
the apparatus must regain the same energy expended at the beginning of the
cycle on $m$. Therefore, owing to the Planck-Einstein formula $E=h\nu$, the
photon frequency $\nu$ must be the same at points ${\cal A}$ and ${\cal B}$.

To emphasize the above conclusion, consider the cycle in reverse.
The first step now consists of the crane emitting a photon of energy $E'$
(frequency $\nu'$) suitably lower than $mgh$. The original energy $E'$ is such
that when the photon arrives at the beacon, it becomes equal to
$E_b = mgh$ ($>E'$) owing to the standard gravitational redshift (blueshift in
this case). In this way, $E_b$ is what is exactly needed to raise the mass
$m$ to the crane at the height $h$. Then, the mass is released back to the
initial position, and the energy coming from that release ($mgh$) goes into
the crane reservoir. At the end of the cycle, the crane will gain positive
energy ($mgh-E'> 0$) out of nowhere.

\section{Energy does not always have mass}
\label{se4}

Now, we have all the tools to show that energy does not always have mass. With
the following thought experiment, we prove that, for instance, the electric
potential energy of a capacitor does not contribute to the capacitor
(gravitational) mass. 

\begin{figure}[t]
\begin{center}
\includegraphics[width=13cm]{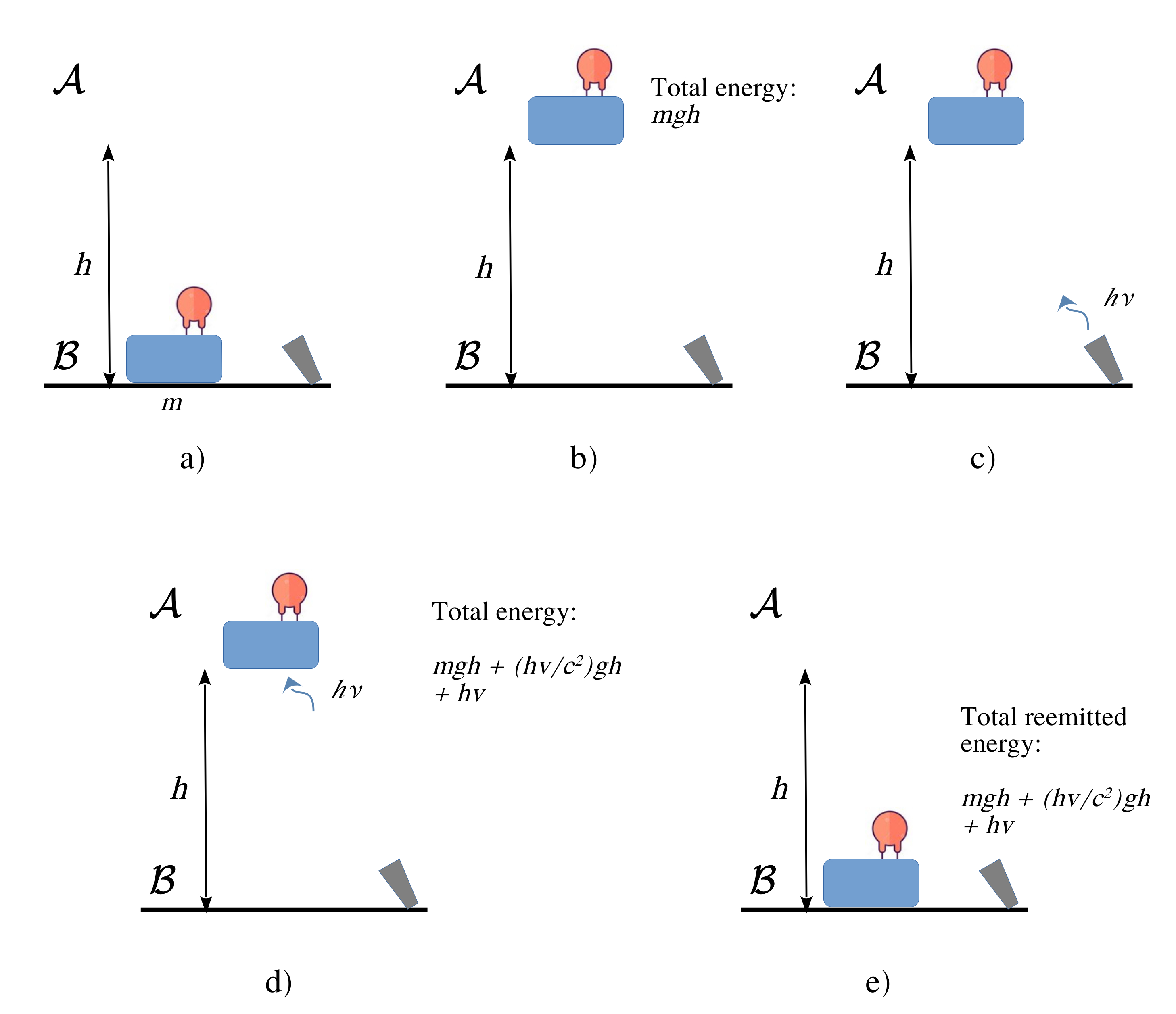}
\end{center}
\caption{Pictorial representation of the thought experiment described in
  Section~\ref{se4}. {\bf (a)} apparatus of mass $m$ initially standing at
  point ${\cal B}$ in a uniform gravitational field $g$; {\bf (b)} raising of
  the apparatus from point ${\cal B}$ to point ${\cal A}$ at a height $h$
  above ${\cal A}$; {\bf (c)} photon emission from the beacon at point
  ${\cal B}$ towards the apparatus at point ${\cal A}$; {\bf (d)} storing of
  the photon energy in the capacitor as electric potential energy; {\bf (e)}
  lowering of the apparatus and capacitor discharging.}
\label{fig4}
\end{figure}

As in Section~\ref{se3}, consider an apparatus of mass $m$ initially standing
at point ${\cal B}$ in a uniform gravitational field $g$ (see Fig.~\ref{fig4}).
This time, the apparatus can convert the incoming radiation energy into
electric potential energy inside a capacitor. The first step of the cyclic
process to be shown consists in raising the apparatus from point ${\cal B}$ to
point ${\cal A}$ at a height $h$ above ${\cal A}$. The work done on $m$ is
equal to $mgh$, which also corresponds to the gravitational potential energy
of the apparatus at point ${\cal A}$. Then, a photon of energy $h\nu$ is
emitted from a beacon at point ${\cal B}$ towards the apparatus at point
${\cal A}$. As established in Section~\ref{se2}, that energy must not change
in climbing up the gravitational field, and, upon absorption by the apparatus,
it is stored in a capacitor as electric potential energy of the same value
$h\nu$.

Now, if the widely-held interpretation that every energy always has mass is
correct, then, upon absorption, the apparatus gains a mass equal to
$\frac{h\nu}{c^2}$. Therefore, the total energy of the apparatus becomes

\begin{equation}
  E_{tot}=mgh + \frac{h\nu}{c^2}gh + h\nu,
\label{eq1}  
\end{equation}  
where $mgh$ is the gravitational potential energy of the apparatus,
$\frac{h\nu}{c^2}gh$ is the gravitational potential energy of mass $h\nu/c^2$,
and $h\nu$ is the energy of the charged capacitor.

As soon as the cycle is completed by lowering the apparatus and discharging the
capacitor, the total re-emitted energy $E_{out}$ needs to be equal to that
given by equation~(\ref{eq1}). That is required by the conservation of total
energy. The problem should now be evident. The input energy $E_{in}$ throughout
the whole cycle is $E_{in}=mgh +h\nu$ while the output energy is
$E_{out}=mgh+\frac{h\nu}{c^2}gh + h\nu$: we have gained an extra-energy
$\frac{h\nu}{c^2}gh$ out of nowhere.

The only possibility to resolve this paradox in compliance with the principle
of conservation of energy is to accept that the energy $h\nu$ stored as
electric potential energy in the capacitor does not have gravitational mass.

There remains one thing to notice. If we do not want to contradict the
conservation of linear momentum, the energy stored in the charged capacitor
has no gravitational mass but must still have linear momentum.
If, like in the thought experiment in Section~\ref{se2}, the apparatus and
the capacitor move horizontally at a constant velocity $v$, the charged
capacitor must have an additional linear momentum equal to
$\frac{{\cal E}}{c^2}v$, where ${\cal E}$ is the electric potential energy
in the capacitor (see also Singal~\cite{sin}). That is not strange. There is
another well-known electromagnetic phenomenon that has momentum but no (rest)
mass: that is light.

\section{Conclusions}
\label{se5}

In the present paper, we have introduced a few thought experiments showing
that energy does not always have mass. For instance, when (radiation) energy is
stored in a reusable form, e.g., the electric potential energy of a capacitor,
that energy does not contribute to the gravitational mass of the device
storing it while still contributing to its linear momentum.
We acknowledge that such a result has fundamental consequences for physics as
we know it (e.g., it might have an impact on the validity of the equivalence
principle), but the derivation is too straightforward to ignore. Moreover, to
this author, our results seem to answer a puzzle relative to a sort of
`doubling of energy'.
For example, if radiation energy is transformed into and stored under the form
of (capacitor) electric potential energy, why should it become mass too?
Isn't mass a further way to store the same energy already stored (and ready
to use) as electric potential energy? To this author, this always appeared
to be a `doubling of energy'.

\section*{Acknowledgments}
The author is indebted to Nils Erik Bomark, Guilherme de
Berredo-Peixoto, Andrea Erdas, Gabriel Ferrari, Daniele Funaro, Espen Gaarder
Haug, Grit Kalies, Ken Krechmer, Peter F.~Lang, Nancy Cambr\'on Mu\~{n}oz,
Bernard Ricardo, Gianfranco Spavieri, and Enayatolah Yazdankish for
stimulating and fruitful discussions on an early draft of the manuscript.

\section*{Data availability}
No new data were created or analysed in this study.

\section*{Competing Interests}
The authors has no relevant financial or non-financial interests to disclose.

\section*{Funding}
The author declares that no funds, grants, or other support were received
during the preparation of this manuscript.

\end{document}